# Dominance of eclipsed ferrocene conformer in solutions revealed by the IR fingerprint spectral splitting

Narges Mohammadi[a], Feng Wang[a*], Stephen Best[b], Dominique Appadoo[c] and Christopher T. Chantler[d]

***(1 June 2013)***

[a]eChemistry Laboratory, Faculty of Life and Social Sciences, Swinburne University of Technology, Hawthorn, Melbourne, Victoria, 3122, Australia
[b]School of Chemistry, University of Melbourne, Parkville, Victoria 3052, Australia,
[c]Far-IR and High Resolution FTIR Beamline, Australian Synchrotron, 800 Blackburn Road, Clayton, Vic. 3168, Australia
[d]School of Physics, University of Melbourne, Parkville, Victoria 3052, Australia

* Corresponding author. Tel.: +61 3 9214 5065; fax: +61-3-9214-5921.

E-mail addresses: fwang@swin.edu.au (F. Wang).

**Abstract**

A combined high-resolution Fourier transform infrared (FTIR) spectra of ferrocene (Fc) and density functional theory (DFT) based quantum mechanical calculations confirmed the dominance of the eclipsed Fc conformer in the fingerprint region of 400-500 $cm^{-1}$. The IR spectra of Fc were measured in solutions with a number of non-polar solvents such as acetonitrile (ACN, ε=35.69), dichloromethane (DCM, ε=8.93), tetrahydrofuran (THF, ε=7.43) and dioxane (DOX, ε=2.21). The measurements agree well with the earlier IR spectra of Lippincott and Nelson (1958) as well as the most recent IR spectral measurement in dichloromethane solution of Duhovic and Diaconescu (2013). All experimental measurements in the solutions unambiguously exhibit an IR spectral splitting of ca. 15 $cm^{-1}$ in the 480-500 $cm^{-1}$ region. The DFT based B3LYP/m6-31G(d) quantum mechanical calculations using implicit solvent models in this study indicates that only the ground electronic state of the eclipsed ($D_{5h}$) Fc splits in the IR fingerprint region of ca. 500 $cm^{-1}$. The IR spectral splitting characterises the centre Fe metal related vibrations of the eclipsed Fc, in agreement with our previous finding in gas phase [Mohammadi et al, 2012]. The present study further suggests that the effects of solvents on the IR spectra of Fc in this region are small and the solvent model effects are also small but the solute molecular density (SMD) model seems to produce the most accurate IR spectrum in the region of 400-600 $cm^{-1}$ of Fc without scaling the calculated results.

## 1. Introduction

Since its discovery some fifty years ago [1], di-cyclopentadienyle iron ($FeCp_2$ or Fc) continues to stimulate an immense number of studies in its own right and in its relation to the prodigious number of cyclopentadienyl (Cp) and other symmetrically delocalized hydrocarbon metal complexes with strong metal-ring π bonding. Exploitation of its well-defined redox behaviour, and ease of derivatization [2]. Have resulted in a broad range of Fc derivatives which have found key roles in homogeneous catalysis, polymer chemistry, molecular sensing, and nonlinear optical materials [3], dye sensitized solar cells (DSSC), as well as biological applications [4, 5].

The ground-state conformation of the two parallel cyclopentadienyl rings of Fc, eclipsed ($D_{5h}$ symmetry) or staggered ($D_{5d}$ symmetry) is surprisingly difficult to resolve unambiguously, this being due to the small difference in energy of the two forms and the low energy barrier for the rotation of a cyclopentadienyl ring relative to the rest of the molecule [6, 7]. This small rational energy barrier is manifest in the experimental studies of Fc since its discovery in 1951 [1]. The debate on the most stable conformer of Fc, whether it is the eclipsed ($D_{5h}$) or the staggered ($D_{5d}$), with both experimental and other information [8-10], has been continuing without conclusive evidences until our recent study of Fc [7]. A few years after the discovery, the infrared (IR) measurements of Fc of Lippincott and Nelson [11-13] were reported.

Despite the accuracy of the early measurements, the IR spectra of Fc were assigned to the staggered ($D_{5d}$) Fc conformer [11], that being the stereochemical form of the complex believed to be present in the crystal structure. This experiment has been considered as providing the assignment of the fundamental vibrational modes for ferrocene [14], however it is important to note that those assignments apply equally to staggered ($D_{5d}$) [11] and eclipsed ($D_{5h}$) Fc conformers.

Detailed structural understanding of the Fc conformers is very important as Fc derivatives may inherit particular properties which only exist in a particular conformer [14, 18]. For example, additional ligands coordinating to the metal and the Cp rings while maintaining certain symmetry is preferred for the geometry of the $D_{5h}$ conformer [14, 19, 20]. Design of synthesis pathways and understanding of the mechanics and reaction dynamics of the Fc derivatives require detailed information of the structure, symmetry and properties of the Fc conformers. The stability of eclipsed and staggered conformers of Fc has been a challenge issue and both structures were discussed in textbooks [15-17]. Recent articles such as Coriani et al [21], Roy et al [22], Gryaznova et al [20] and Bean et al [23] have well documented the history and current status of the Fc studies.

Properties of the staggered and eclipsed conformers of Fc are not markedly different due to the shape of the conformers. The differences between the calculated properties of the Fc conformers are either too small to measure or too small to differentiate (e.g., within the error of the methods), such as dipole moment and rotational constants [7]. This explains why previous conclusions regarding the conformation of Fc are contradictory [8-10]. This statement being able to be tested by the calculations of properties such as the IR spectra [7, 11] and it is timely to review the assignment of the observed IR spectra to the staggered ($D_{5d}$) conformer of Fc [11]. Recent infrared (IR) spectral simulations of Fc [14, 20] revealed that the DFT methods can be employed to interpret the measured IR spectra of Fc [11]. However, the basis sets employed in the DFT models [14, 20] were not particularly suitable to the iron atom of Fc, the simulated IR spectrum of Fc needed scaling factors to fit the measurements [14, 20]. As a result, the simulated Fc IR spectra may be useful for interpretation but hardly suitable for prediction purposes. Whether the measured IR spectrum was the staggered Fc, or the eclipsed Fc or a mixture of both, still remains an open question in the study of Fc.

In our recent DFT based study [7], it was proposed that the spectral splitting within the IR region of 480-500 $cm^{-1}$ provides a means of distinguishing between the eclipsed and staggered Fc conformers. It was further demonstrated that the vibrations of Fc in this region are dominated by the vibrations of the centre Fe atom of the sandwich complex, which represents the major conformational differences between the eclipsed and staggered conformers [7]. As a result, the related IR spectral peak in the 480-500 $cm^{-1}$ region splits if it is an eclipsed orientation; otherwise it is the staggered conformer of Fc and there is a significant difference in wavenumber of the bands for the staggered and eclipsed forms. Therefore it should be possible to identify from simple IR measurements whether there is a single conformer or a mixture of the two forms. Even though in high resolution FTIR spectra of Fc, this IR peak of the staggered conformer can hardly split as the vibrations of the $D_{5d}$ conformer were predicted within only approximately 2 $cm^{-1}$ in gas phase [7]. The DFT based B3LYP/m6-31G(d) model accurately simulated the IR spectra of Fc ***without scaling***, as the basis set [24] for the central Fe atom of the sandwich compound plays a significant role [7]. For the Fe atom, the modified 6-31G(d) basis set, i.e., m6-31G(d) [24] is so far the most appropriate basis set for accurately modelling the complexes containing Fe.

In the present study, we provide a combined study of DFT calculations with high-resolution FTIR experimental measurements in a number of solvents, in order to validate the earlier theoretical prediction of the Fc conformers [7].

## 2. Experimental details and computational methods

Infrared (IR) spectra of Fc dissolved in non-polar solvents such as acetonitrile (ACN, $\varepsilon$=35.69), dichloromethane (DCM, $\varepsilon$=8.93), tetrahydrofuran (THF, $\varepsilon$=7.43) and dioxane (DOX, $\varepsilon$=2.21) were obtained using a Bruker Tensor 27 FTIR and a conventional solution cell fitted with KBr windows and a 100 μm spacer. A resolution of 1 $cm^{-1}$ was used for all measurements. Concentrated solutions were prepared in each case and these spectra were compared with those obtained from the corresponding solutions diluted by factors ranging between 2 and 4. No significant concentration-dependent variation in the band profile was observed.

The optimized geometries of Fc eclipsed and staggered conformers were obtained using the B3LYP/m6-31G(d) model [24]. The DFT model is the same model in our previous study of Fc in gas phase [7]. This m6-31G(d) basis set incorporates necessary diffuse d-type functions for the Fe transition metal, so that it ensures a better performance than the conventional 6-31G(d) basis set for the iron atom of Fc, by providing a more appropriate description for the important energy differences between the atomic $3d^n4s^1$ and $3d^{n-1}4s^2$ configurations [25]. The present IR spectral simulations of the Fc conformers use the same model of B3LYP/m6-31G(d) in the solvents ***without any scaling***.

As the IR spectra of Fc were measured in solutions, a number of implicit solvent models were employed in the simulations. Continuum solvation methods such as polarizable continuum model (PCM) or dielectric polarized continuum model (D-PCM which neglects the volume polarization) [26], the conductor PCM (C-PCM, which approximates the volume polarization) [27, 28] and the solute molecule density (SMD) model [29] were employed in the IR spectral simulations. Geometry optimization calculations of Fc were performed for each such calculation, respectively, followed by vibrational frequency calculations. All calculations are performed using the Gaussian09 computational chemistry package [30].

## 3. Results and discussion

Figure 1 shows the analysis of the Fc IR spectrum in the acetonitrile (ACN, ε=35.69) solution in the region of 400-1200 $cm^{-1}$. Five major spectral peaks (four bands) are measured in this region as given in Figure 1(a). The four measured major peaks (five bands) of Fc are analysed (Voigt fits) at 480.28±0.05 $cm^{-1}$ (Int: 0.12±0.001) and 496.25±0.03 $cm^{-1}$ (Int: 0.19±0.001) in Figure 1 ($b_1$), 823.69±0.03 $cm^{-1}$ (Int: 0.20±0.002) in Figure 1 ($b_2$), at 1006.60±0.02 $cm^{-1}$ (Int: 0.16±0.001) in Figure 1 ($b_3$), and 1107.60±0.02 $cm^{-1}$ (Int: 0.25±0.002) in Figure 1 ($b_4$). Table 1 summarised the FTIR spectral information, including positions, height of intensities and full width at half maximum (FWHM) etc. The intensity ratio of the measured major spectral peaks in the ACN solution from 400-1200 $cm^{-1}$ is 1:1.58:1.71:1.38:2.14, which is in good agreement with the recent IR measurement of Fc in DCM solution which is given by 1 : 1.39 : 1.53 : 1.33 : 1.55 [14].

Figure 2 compares the FTIR spectral region of 400-1200 $cm^{-1}$ in solvents at room temperature. The FTIR spectra were recorded non-polar solvents such as acetonitrile (ACN, ε=35.69), dichloromethane (DCM, ε=8.93) and tetrahydrofuran (THF, ε=7.43). As seen in this figure, the spectra all exhibit four major peaks at ca. 480-500 $cm^{-1}$, 820 $cm^{-1}$, 1010 $cm^{-1}$ and 1100 $cm^{-1}$. The present FTIR spectra in solutions consistently show excellent agreement with the Fc IR spectrum of Duhovic and Diaconescu in the DCM solution [14] as well as the early IR spectrum of Lippincott and Nelson [11] in chloroform solution ($CCl_4$) at room temperature.

Table 2 reports the Fc IR spectral peak positions in the solutions and compares with other experiment measurements and calculations in 400-1200 $cm^{-1}$. The present experimental measurements in different non-polar solvents consistently agree with available IR measurements in solutions in this region. For example, the IR spectral peak positions in the dichloromethane (DCM) solution were measured as 478 $cm^{-1}$ and 494 $cm^{-1}$, 822 $cm^{-1}$, 1004 $cm^{-1}$ and 1106 $cm^{-1}$ in a recent study [14], whereas the present measurement reports as 478 $cm^{-1}$ and 495 $cm^{-1}$, 820 $cm^{-1}$, 1005 $cm^{-1}$ and 1107 $cm^{-1}$, accordingly, almost exactly reproduce all IR spectral peak positions in the region of 400-1200 $cm^{-1}$. In addition, our experimental results show that the positions of the spectral peaks the non-polar solvents are almost identical, which are consistently reported within ±3 $cm^{-1}$ of accuracy, which indicates that solvent effects to the IR spectra of Fc are small and negligible.

To further obtain theoretical insight of the Fc structural information from the FTIR measurement, Figure 3 compares the measured FTIR spectrum (middle panel) of Fc in the acetonitrile (ACN) solution with the simulated infrared spectra of the eclipsed ($D_{5h}$, top panel) and the staggered ($D_{5d}$, bottom panel) conformers of ferrocene in the region of 400-1200 $cm^{-1}$. The simulated IR spectra were obtained using the B3LYP/m6-31G(d) model ***without scaling and shifting***. In the simulation, the implicit solute molecular density (SMD) model [29] is applied. A general agreement between the measurement and simulation is achieved. For example, in both Fc conformers, the major FTIR spectral peaks in the measurement (middle panel) are reproduced by the simulations. For example, the measured peak positions are given at 480 $cm^{-1}$/496 $cm^{-1}$, 823 $cm^{-1}$, 1006 $cm^{-1}$ and 1107 $cm^{-1}$ in the ACN solution. In the simulated IR spectrum of the eclipsed Fc conformer in the same solution, these spectral peaks are given as 467 $cm^{-1}$/484 $cm^{-1}$, 852 $cm^{-1}$, 1027 $cm^{-1}$ and 1133 $cm^{-1}$, accordingly, without any scaling and shifting. These IR peaks of the staggered Fc conformer, on the other hand, are given by 458 $cm^{-1}$ (only one peak), 854 $cm^{-1}$, 1029 $cm^{-1}$ and 1134 $cm^{-1}$, accordingly, under the same conditions.

Combining the experimental results with theory, Figure 3 reveals that only the IR spectral peak(s) of the eclipsed Fc conformer in the region under 500 $cm^{-1}$ splits into two peaks, a more intensive peak at a larger frequency (i.e. 495 $cm^{-1}$) and a less intensive peak at a smaller frequency (i.e. 479 $cm^{-1}$). The staggered Fc conformer, however, does not show such the splitting, in agreement with our previous study in gas phase [7]. Such the IR spectral splitting of Fc has been observed by a number of earlier experiments including Lippincott and Nelson [11] and Duhovic and Diaconescu [14]. This particular spectral feature in the IR fingerprint region of 400-500 $cm^{-1}$ is considered as the signature of the eclipsed Fc conformer. It is noted in Figure 3 that the calculated spectral peaks red-shift (i.e. smaller than the measured) in the region of below 500 $cm^{-1}$ but blue shift (i.e. larger than the measured) in the region of above 500 $cm^{-1}$ from the measurements. Further investigation into the related vibrations reveals that the spectral peak(s) in the region below 500 $cm^{-1}$ are associated with Fe-C stretches, whereas the vibrations at 845 $cm^{-1}$, 1032 $cm^{-1}$ and 1141 $cm^{-1}$ are assigned to C-H out of plane, C-C-H bend and Cp breathing (one Cp shrinks, one Cp expands) vibrations, respectively. This is in agreement with previous studies of ferrocene [7, 11, 20].

Previous studies [20, 31, 32] indicated that the scaling factors of simulated IR spectra (force field) of compounds are usually smaller than 1.0 [33-35]. That is, the calculated vibrational frequencies are usually larger than the measurement, except for the Fe-C stretch vibration which needs a scaling factor as large as 1.25 [20] (1.02 in the present study due to the more suitable m6-31G(d) basis set)---the calculated frequencies of Fe-C related vibrations are smaller than the measured ones. Table 3 compares the measurement and calculations for the spectral peaks of the eclipsed Fc in different solutions. The largest error is within 4% between the measurements and calculations in the region of 400-1200 $cm^{-1}$, indicating that the theoretical model is sufficiently accurate for the Fc conformers. The red shift (by colour) for the Fe related vibrations and the blue shift for other vibrations are apparently shown in this table.

The present study will concentrate on IR spectra of the eclipsed Fc in the region of 400-500 $cm^{-1}$, as the most noticeable feature of the FTIR spectra in Figure 2 and Figure 3 is the spectral peak splitting in the region of approximately 480-500 $cm^{-1}$ in all solutions. The major peak locates at 495 $cm^{-1}$ whereas the minor peak locates at 479 $cm^{-1}$ (note that the resolution of the Bruker Tensor FTIR spectrometer is 1 $cm^{-1}$), with a measured spectral splitting of approximately 16 $cm^{-1}$ in the ACN solution. The splitting of the twin peak agrees well with the earlier IR measurement of Lippincott and Nelson [11], in which the spectral peaks located at 492 $cm^{-1}$ and 478 $cm^{-1}$ in chloroform solution with a spectral splitting of 14 $cm^{-1}$, and with the recent FTIR measurement at 478 $cm^{-1}$ and 494 $cm^{-1}$ in DCM solution with a splitting of 16 $cm^{-1}$ of Duhovic and Diaconescu [14]. Note that the present IR calculations do not use any scaling whereas the calculated IR spectra of Ref [14] employed a scaling factor of 0.95.

Ferrocene resolves in non-polar solvents as both the eclipsed and staggered conformers lack a permanent dipole moment due to the highly symmetric di-cyclopentadienyle structure. In Figure 4, the FTIR measurements in the solutions are presented in high and low concentrations in the IR region of 400-600 $cm^{-1}$. The FTIR spectra do not exhibit apparent solvent dependent changes nor concentration dependent changes. In particular, the spectral peak positions in the solvents, such as DCM, THF and DOX which have small dielectric constants (i.e., $\varepsilon<10$), are almost identical. The figure also shows that the concentration of the solutions do not change the positions of the spectral peaks, except for the intensities. That is, the intensive red spectra in Figure 4 are in solutions with high concentration and the less intensive light blue spectra represent solutions with low concentration. No apparent concentration related spectral shifts are observed in the measured FTIR spectra, suggesting small and negligible effects of the concentration of the solutions on the IR spectra of Fc.

Of the implicit solvent models, some are able to reproduce the experiment measurements better than other solvent models. Figure 5 compares the simulated IR spectra of eclipsed Fc in dioxane solution using three different models including polarizable continuum model (PCM) [26], C-PCM (conductor PCM) [27, 28] and SMD (solute molecule density) [29], with the present FTIR spectral measurement in high concentration. The solute molecule density (SMD) model seems to achieve a slightly more accurate agreement with the measurement over the PCM and C-PCM models. The PCM and C-PCM solvent models only implicitly consider the solvation effects in the form of a polarizable continuum model rather than individual molecules. For example, the conventional PCM model (also called dielectric PCM or D-PCM [36] ) treats the solvent as a polarizable dielectric continuum medium. For non-polar solvents the PCM and C-PCM models do not show apparent differences over the IR spectra of Fc. The solute molecule density (SMD) model is a continuum solvation model based on the quantum mechanical charge density of a solute molecule interacting with a continuum description of the solvent. It more accurately simulates the IR spectra of Fc in solutions and therefore, it is employed to simulate the solvent effects on the IR spectra of ferrocene in this study.

Figure 6 compares the simulated IR spectra of the eclipsed Fc in the region of 400-600 $cm^{-1}$ with the FTIR measurement in several solvents. The theory (simulation) used the DFT based B3LYP/m6-31G(d) model in conjunction with the SMD continuum solvent model to simulate the IR spectra of the eclipsed Fc in THF, DCM, ACN and DOX solvents. The theoretical spectra of the eclipsed Fc are produced in the present study ***without any scaling*** but the simulated spectra in Figure 6 are presented with a blue shift of 6 $cm^{-1}$ and 10 $cm^{-1}$ for DOX and THF solvents, respectively, and of 11 $cm^{-1}$ for DCM and ACN solvents, respectively. As seen from this figure, after small shifts, the simulated spectra agree well with the measurements including the twin peak splitting of the eclipsed Fc.

## 4. Conclusions

The present study reveals the dominance of the eclipsed ferrocene conformer in solutions at room temperature. It combines high-resolution Fourier transform infrared (FTIR) spectral measurements with theoretically simulated spectra of ferrocene in a number of non-polar solvents, such as ACN, DCM, THF and DOX, respectively, in a region of 400-1200 $cm^{-1}$. Furthermore, the solutions in high and low concentrations are measured in the region of 400-600 $cm^{-1}$, respectively. The measurements consistently agree well with previously available IR spectra in $CCl_4$ solution of Lippincott and Nelson [11] as well as the most recent IR spectral measurement in dichloromethane (DCM) solution [14]. The IR spectra of ferrocene are also simulated using density functional theory (DFT) based B3LYP/m6-31G(d) model in acetonitrile solution for both eclipsed and staggered Fc conformers. All experimental measurements in solutions unambiguously exhibit an IR spectral splitting. When combined the experimental results with theory, it is concluded that the spectral splitting in the IR fingerprint region of ca. 500 $cm^{-1}$ can only be the structure of eclipsed Fc ($D_{5h}$), whereas this spectral peak of the staggered Fc ($D_{5d}$) do not split. This finding is in agreement with our previous discovery in gas phase [7]. The present study further investigated the effects of solvents on the IR spectra and the solvent model effects on the simulated spectra. It is found that the IR spectra of ferrocene are not apparently solvent dependent. Only small spectral shifts are due to different solvent models but the solute related model, i.e., the solute molecular density (SMD) model seems to produce the most accurate IR spectrum in the region of 400-600 $cm^{-1}$ of ferrocene.

## Acknowledgements

NM acknowledges Vice-Chancellors' Postgraduate Research Award at Swinburne University. FW thanks the National Computational Infrastructure (NCI) at the Australian National University for an award under the Merit Allocation Scheme, Victorian Partnership for Advanced Computing (VPAC) and the Swinburne University Supercomputing for computing facilities. FW, SB and CC acknowledge Australian Synchrotron Research Program which is funded by the Commonwealth of Australia under the Major National Research Facilities Program.

**Table 1:** Voigt fits (Lorentzian-Gaussian) of the FTIR spectra measured of ferrocene in ACN solution.

| **Solvent=ACN** | $\upsilon_1$ | $\upsilon_2$ | $\upsilon_3$ | $\upsilon_4$ | $\upsilon_5$ |
|---|---|---|---|---|---|
| **Position** | 480.28 | 496.25 | 823.69 | 1006.60 | 1107.60 |
| | ±0.05 | ±0.03 | ±0.03 | ±0.02 | ±0.02 |
| **Intensity** | 0.12 | 0.19 | 0.20 | 0.16 | 0.25 |
| | ±0.001 | ±0.001 | ±0.002 | ±0.001 | ±0.002 |
| **FWHW** | 8.91 | 8.09 | 11.79 | 7.67 | 3.21 |
| **GL** | 0.47 | 0.27 | 0.66 | 0.8 | 0.52 |

**Table 2:** Comparison of the measured Fc spectral peak positions in various solvents and available experiment and calculations.

| | This Work[a] | | | | | | Ref.[14] | | Ref.[20] | Ref.[11] | |
|---|---|---|---|---|---|---|---|---|---|---|---|
| | Expt. | | | | B3LYP/m6-31G(d) | | Expt. | B3LYP/ 6-31G*[b] | B3LYP/ Type-I | Expt. | |
| | ACN | DCM | THF | DOX | DOX[a] | Gas[7] | DCM | Gas | Gas | $CCl_4$ | Gas |
| $\upsilon_1$ | 480 | 478 | 479 | 479 | 471 | 471 | 478 | 480 | 470 | 478 | 480 |
| $\upsilon_2$ | 496 | 495 | 495 | 495 | 489 | 488 | 494 | 510 | 473 | 492 | 496 |
| $\upsilon_3$ | 823 | 820 | 820 | 821 | 845 | 844 | 822 | 844 | 825 | 811 | 816 |
| $\upsilon_4$ | 1006 | 1005 | 1005 | 1006 | 1033 | 1035 | 1004 | 1036 | 1017 | 1002 | 1012 |
| $\upsilon_5$ | 1107 | 1107 | 1108 | -[c] | 1141 | 1141 | 1106 | 1140 | 1141 | 1108 | 1112 |

[a] DFT model: SMD-B3LYP/m6-31G(d).

[b]A scaling factor of 0.95 was employed in the calculations in Ref [14] and private communications (2013).

[c]The spectrum cut off in this solution was after 1100 $cm^{-1}$ as the measurement in this solution concentrated in the region under 1000 $cm^{-1}$.

**Table 3:** Comparison of the measured and simulated Fc IR spectral peak positions in solvents in the region of 400-1200 $cm^{-1}$.

| | **ACN** (ε=35.69) | | | **DCM**(ε=8.93) | | | **THF**(ε=7.43) | | | **DOX**(ε=2.21) | | |
|---|---|---|---|---|---|---|---|---|---|---|---|---|
| | **Expt.** | **Calc.**[a] | **Δυ**[b] | **Expt.** | **Calc.**[a] | **Δυ**[b] | **Expt.** | **Calc.**[a] | **Δυ**[b] | **Expt.** | **Calc.**[a] | **Δυ**[b] |
| $\upsilon_1$ | 480 | 467 | -13 | 478 | 466 | -12 | 479 | 467 | -12 | 479 | 471 | -8 |
| $\upsilon_2$ | 496 | 484 | -12 | 495 | 484 | -11 | 495 | 485 | -10 | 495 | 489 | -6 |
| $\upsilon_3$ | 823 | 852 | 29 | 820 | 850 | 30 | 820 | 851 | 31 | 821 | 845 | 24 |
| $\upsilon_4$ | 1006 | 1027 | 21 | 1005 | 1029 | 24 | 1005 | 1029 | 24 | 1006 | 1033 | 27 |
| $\upsilon_5$ | 1107 | 1133 | 26 | 1107 | 1134 | 27 | 1108 | 1134 | 26 | - | 1141 | - |

(a) Simulation DFT model: SMD-B3LYP/m6-31G(d).

(b) Spectral shift: $\Delta\upsilon = \upsilon_{calc.} - \upsilon_{expt}$

## Figure captions

**Figure 1** The measured FTIR spectrum of ferrocene in ACN solution (a) and Voigt fits for 4 bands (5 peaks) in the region of 400-1200 cm $^{-1}$ ($b_1$-$b_4$) .

**Figure 2** Measured FTIR spectra of ferrocene in the region of 400-1200 cm$^{-1}$ in a number of solvents at room temperature.

**Figure 3** The measured IR spectrum (middle panel) of Fc in acetonitrile (ACN) solution with the simulated infrared spectra of the eclipsed ($D_{5h}$, top panel) and the staggered ($D_{5d}$, bottom panel) conformers of Fc in the region of 400-1200 cm$^{-1}$. The spectrum clearly indicates the dominance of eclipsed Fc in this solution.

**Figure 4** The measured FTIR spectra in the solutions of high and low concentrations in the IR fingerprint region of 400-600 cm$^{-1}$.

**Figure 5** The simulated IR spectra of eclipsed Fc in the DOX solution using three different models (PCM, CPCM and SMD) with the FTIR spectral measurement in high concentration. The SMD model show slightly more accurate spectrum.

**Figure 6** Comparison of the simulated IR spectra of the eclipsed Fc in the region of 400-600 cm$^{-1}$ with the FTIR measurement in various solvents. Small shift to align the larger peak is applied.

Fig 1

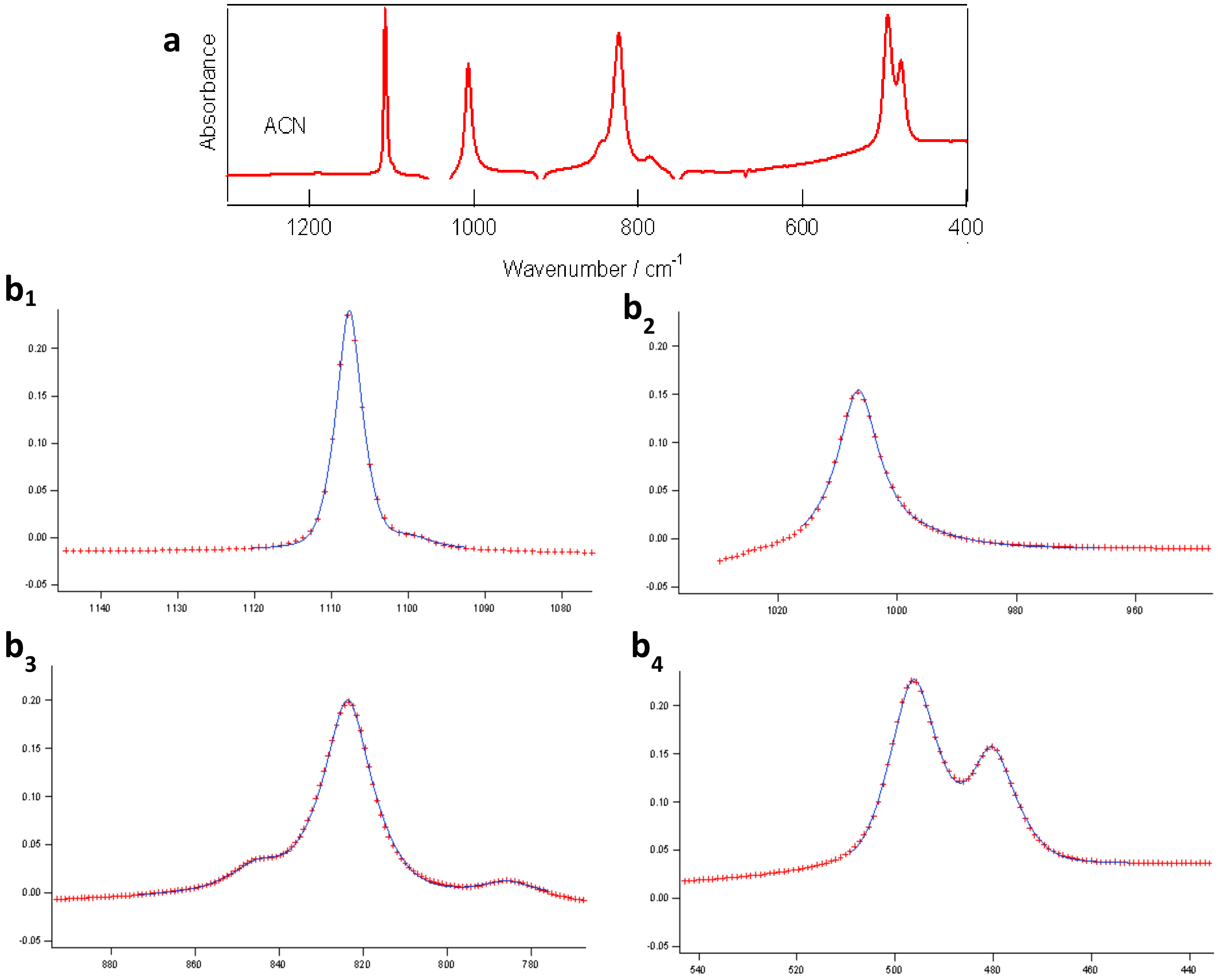

Fig2

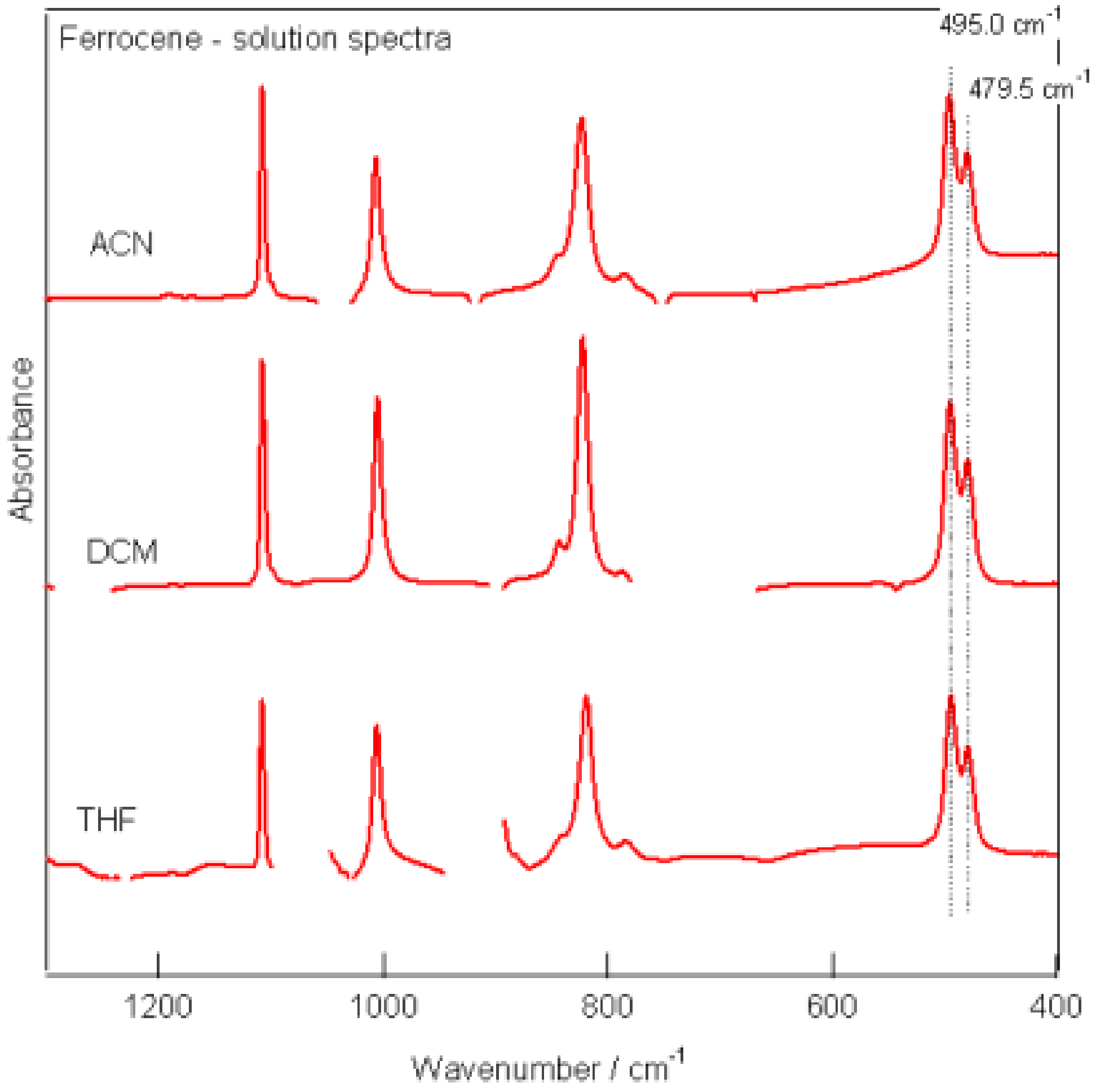

Fig3

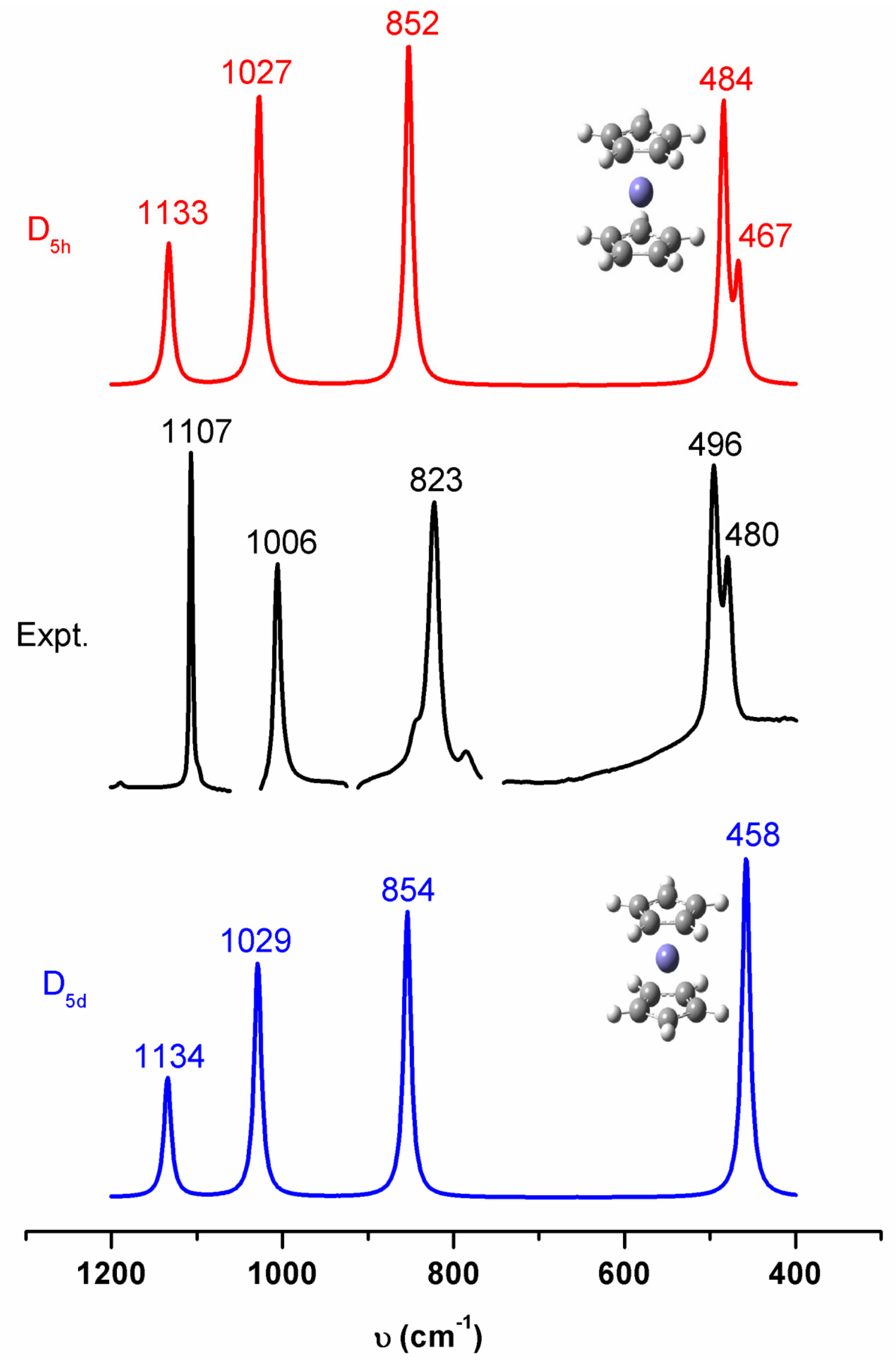

Fig4

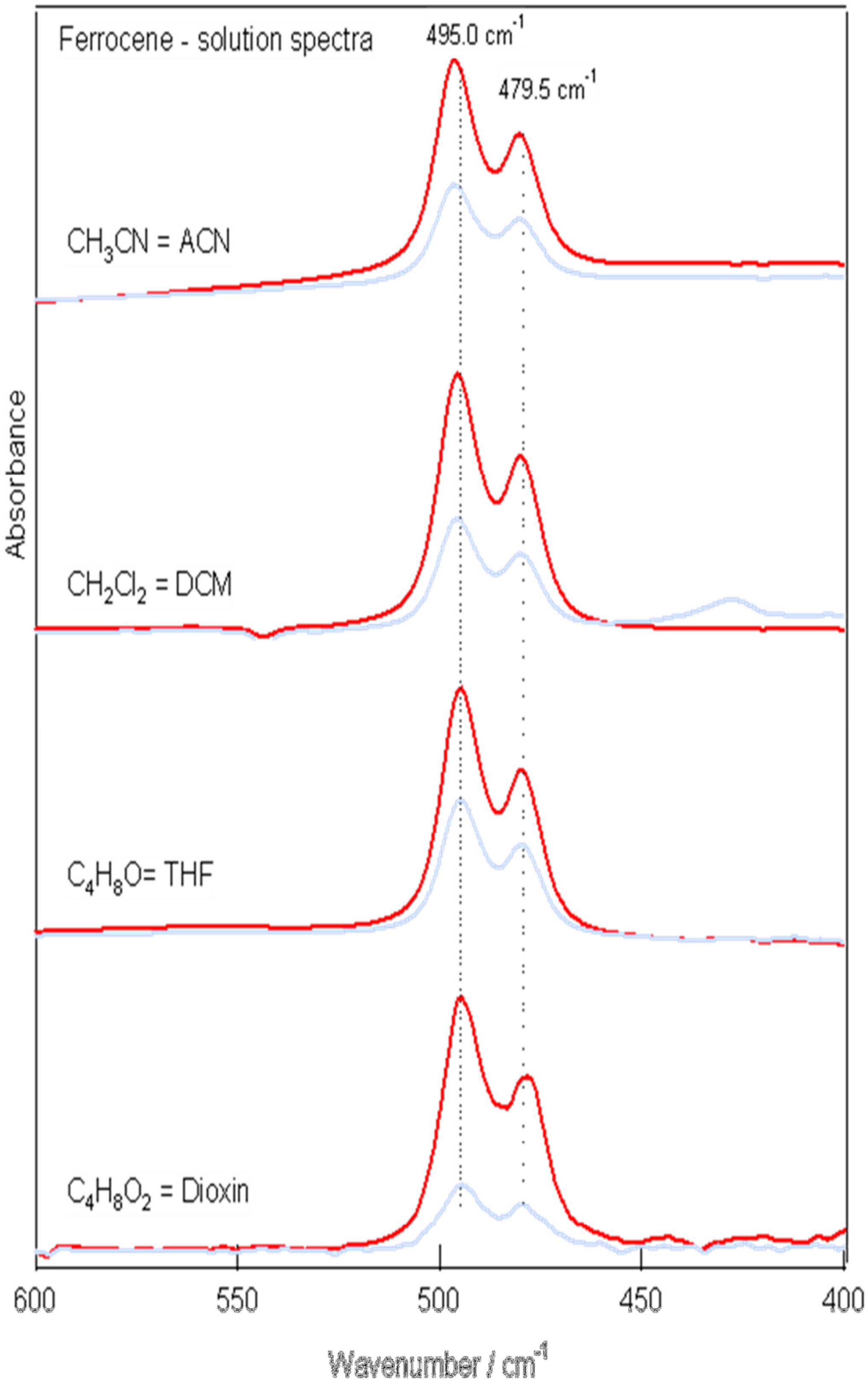

Fig5

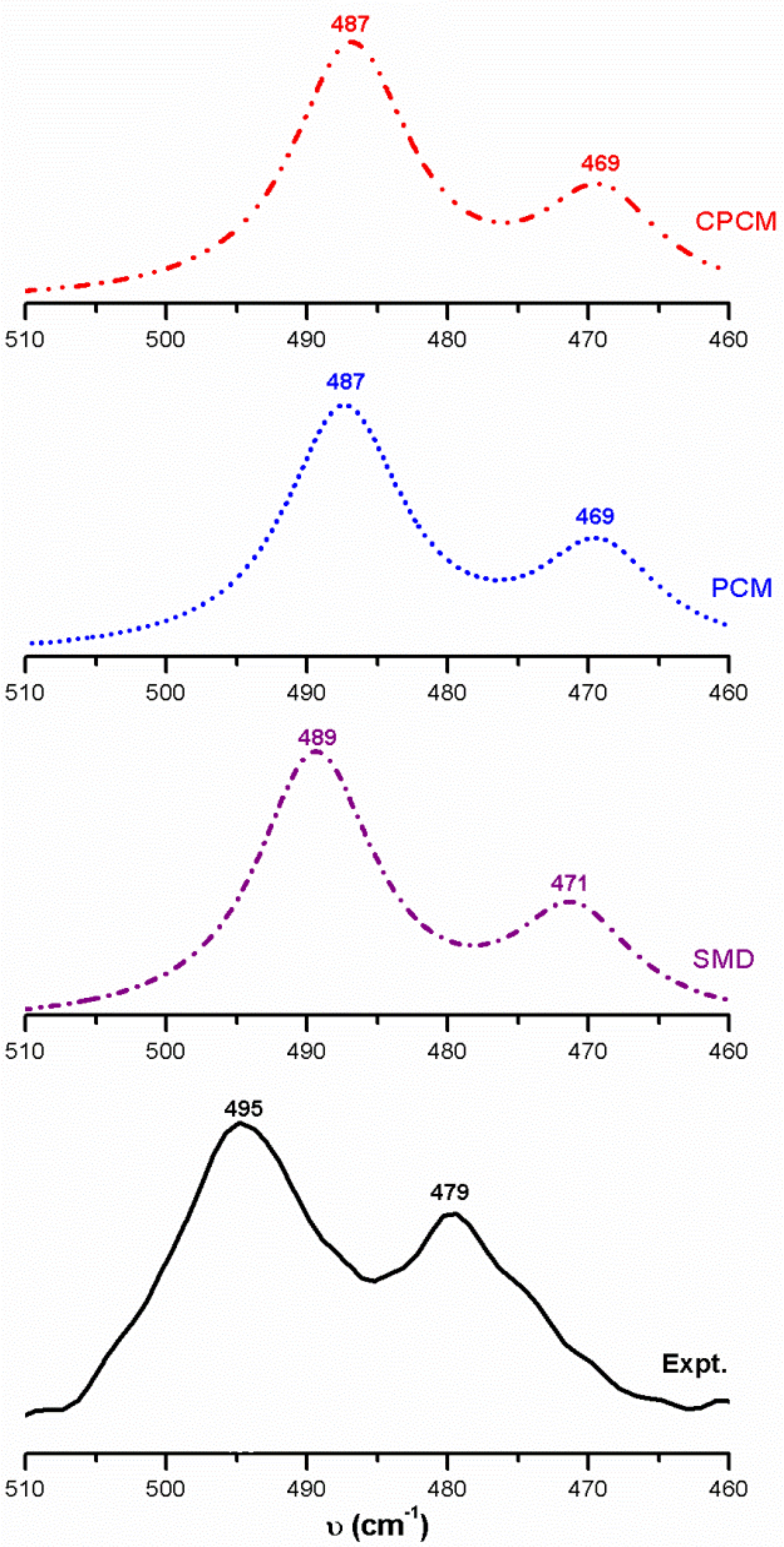

Fig6

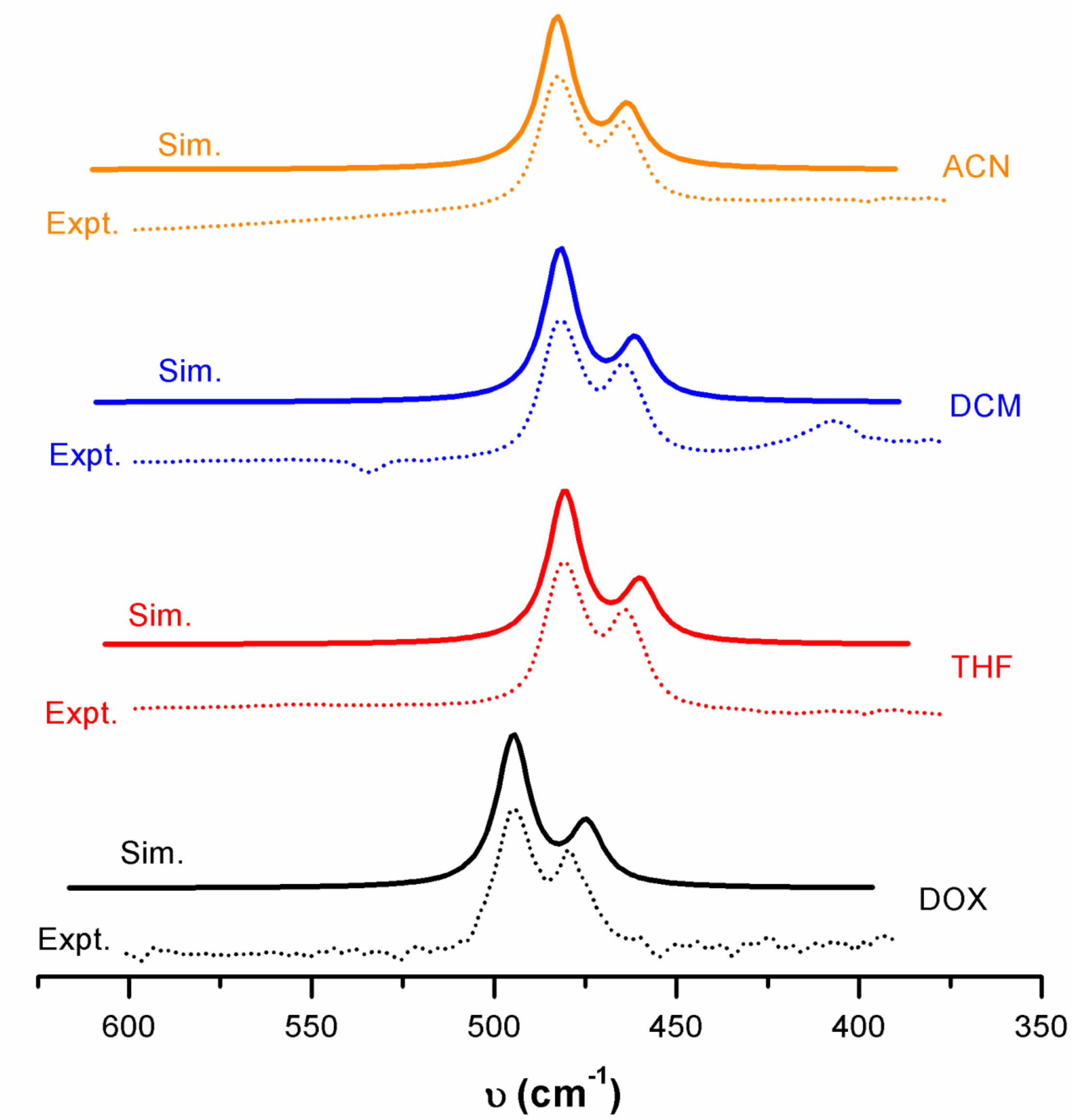